\documentclass[preprint]{aastex}

\usepackage{graphicx}

\usepackage{natbib}

\newcommand{\be}{\begin{equation}}
\newcommand{\ee}{\end{equation}}
\newcommand{\nn}{\mbox{} \nonumber \\ \mbox{} }
\newcommand{\ba}{\begin{eqnarray}}
\newcommand{\ea}{\end{eqnarray}}
\newcommand{\om}{\omega}

\newcommand\etal{\textit{et al.}}
\newcommand\eg{\textit{e.g.}}

\newcommand{\Bf}{{magnetic field}}

\newcommand{\NS}{neutron star}

\begin{document}

\title{Double explosions and jet formation in GRB-supernova progenitors}

\author{Maxim Lyutikov\\
Department of Physics, Purdue University, \\
 525 Northwestern Avenue,
West Lafayette, IN
47907-2036 }

%\date{Received   / Accepted  }

\begin{abstract}
Progenitors of long GRBs, and core-collapse supernovae in general,  may   have two separate  mechanisms driving the outflows: quasi-isotropic neutrino-driven  supernova explosions followed by a highly collimated relativistic outflow driven by the GRB central engine, a black hole or a magnetar. We consider the dynamics of the second GRB-driven explosion
 propagating through expanding envelope generated by the passage of the primary supernova shock. Beyond the central core, in the region of steep density gradient
 created by the SN shock breakout, the 
accelerating  secondary quasi-spherical GRB shock  become unstable
to corrugation and under certain conditions may form a highly collimated jet, a ``chimney'',  when a flow expands almost exclusively along a nearly cylindrically collimated channel.  Thus, weakly non-spherical driving and/or non-spherical initial conditions of the wind cavity
may  produce highly non-spherical, jetted outflows. For a constant luminosity GRB  central engine,  this occurs for density gradient  in the envelope $\rho \propto r^{-\om}$ steeper than $\om >4$.

\end{abstract}
\maketitle

\section{Double explosion in core collapse supernovae}

Long Gamma Ray Bursts (GRBs) are intrinsically linked to core collapse supernovae. This conclusion comes from the 
detection 
of Type Ic supernovae nearly coincident with long GRBs  \citep{smg+03,Hjorth}. It is also confirmed by 
 studies of the host galaxies 
 of long GRBs, which turned out to be actively star-forming  \citep{Djorgovski}.
 The leading model of  long GRBs is a collapsar model \citep{1999ApJ...524..262M,2008MNRAS.385L..28B}, which postulates that a compact central source (a black hole or rapidly rotating \NS\ \cite{Usov92}) forms inside the collapsing  core. The central engine generates a collimated  outflow, which upon breaking out of  the star reaches relativistic velocities and eventually produces $\gamma$-rays.
 
Modern models  of neutrino-driven SN explosion are not 
ÒstableÓ, in a sense that  different groups do not 
agree with each other and the role of different ingredient is not settled \citep[\eg][]{2009AIPC.1171..273B}. 
The collapsar model assumes that in addition to the conventional neutrino-driven SN explosion, there is an addition source of energy, the GRB central engine. 
%In addition, the role of \Bfs\ is not clear  \citep[the idea of magnetically-driven explosions dates back to][]{LeBlancWilson,Bisnovatyi71}: in {\it all}  present-day  simulations of a  successful magnetic 
%explosions, a  very high 
%initial \Bf\  is needed \citep{2007ApJ...664..416B,2009arXiv0907.0561N}.
% It is feasible that such field are indeed achieved in nature, but testing this paradigm requires full 3-D relativistic MHD simulations of magnetic dynamo; something  which has not been done yet for this problem.  Thus, the
%initial conditions for the GRB engine are not clear. 
It is possible that depending on the detailed properties  of the pre-collapse core (like angular momentum, initial \Bf, small differences in composition etc), the
two energy sources  that may potentially lead to the explosion, neutrino-driven convection and the GRB central engine, may contribute different amount of energy, resulting in different observed types of SNe and/or GRBs. Neutrino-driven explosions generates  quasi-spherical sub-relativistic outflows, while GRB explosion results in a  jetted relativistic component. Classical SNe are all neutrino driven, where  GRB engine is  negligible. 
As the  relative strength of the GRB engine increases, this leads to phenomena of sub-energetic and regular GRBs. The relative contribution of the 
neutrino-driven and GRB energies are definitely not independent of each other: one expects that in case of a  successful neutrino-driven SN, the amount of material accreted on the central source is small, resulting in weak or no relativistic component and a weak GRB:  recall that all SN-associated GRBs are subluminous. 

Thus, a SN explosion can be viewed as a two-parameter phenomenon, the two parameters being the power of the neutrino-driven and GRB-driven outflows. Most supernova are neutrino-driven quasi-spherical outflows, where the GRB-driven component is weak or non-existent.  
The recent  discovery of the  relativistic type Ibc supernova without a detected GRB signal \citep[SN 2009bb][]{2010Natur.463..513S} requires both two
energy sources:  one to generate the relativistic outflow, another to expel the SN envelope. In addition,
some of the well studied  supernova remnants, like Cas A, do show jet-like features. \cite{LamingCasAGRB} indeed suggested that Cas  A could have been a failed GRB, generating a jet with a typical energy order of magnitude smaller than a typical long GRB.

 As the power of the neutrino-driven outflow decrease, the fall-back material may power the GRB-like central engine. Thus, 
  the SN  and GRB explosions  are two related, but different events: SN shock expels or nearly expels the envelope, while the GRB outflow is concentrated in a narrow solid angle, presumably along the axis of rotation of the central object. In this picture, the GRB engine does need (but still may) to contribute to overall dynamics of the  envelope: most of the heavy lifting (unbinding the envelope) is done by the neutrinos. 
  
For the purpose of this paper, we accept this paradigm, that both GRBs and some non-GRB SNe
 have two driving mechanisms with the relative energies of the two outflows varying over a large range.  We assume that GRB shock follows that of the SN, but not by much, so that  the confining ram pressure of the ejecta is  important for propagating of the secondary GRB shock. 
 
   In passing we note that a two stage model of SN explosions have already been advocated by 
 \cite{Grasberg} long before the discovery of jet-like features in SNe. This was based on modeling of line emission, in particular of remarkably narrow emission and absorption lines. 
 They suggest that weak explosion proceeded the SN shock, though. Also, a supranova model of \cite{2002astro.ph.11300D} postulates a  two-staged explosion.  
 %(Also,  the model was developed for SN II type explosions, while GRBs are associated with type Ib/c).
 
  \section{Formation of a GRB jet: Kompaneets approximation}
 
 How is the GRB jet collimated? 
Most  models of jet formation in a collapsing star employ a  highly anisotropic driving, either 
 through  neutrino-induced heating  \cite{1999ApJ...524..262M} or  magnetic collimation (\cite{lb03,KomissarovGRBSN,BuciantGRB}). In this paper we investigate an alternative mechanism that produces a highly anisotropic outflow, while been driven by a {\it weakly}  anisotropic central source. The collimation mechanism relies on the ram pressure of the external mediums and large scale Raleigh-Taylor  (RT) instability of accelerating shocks.

There is an extensive literature on RT instability of {\it point } explosions propagating in steep density gradient
\citep[\eg][]{RyuVishniac,Chevalier90}. The particular case of a driven shock has not been investigated to the best of our knowledge. Another qualitative difference of our approach from the conventional  SN theory \citep{RyuVishniac,Chevalier90} is that we allow much steeper external density gradients, established by a passage of a preceding SN shock.

 The distinction between point explosions propagating in steep density gradient, and  driven explosions  (sub-sonically or supersonically) is important. Accelerating shocks from point explosions obey the so called second type of self-similarity \citep{ZeldovichRaizer}, whereby the flow passes through critical point so that the flow dynamics is, in some sense, self-determined. In subsonically driven explosions the shock obeys a Sedov-type scaling, where post-shock pressure is related to the average pressure in post-sock cavity.  Supersonically driven shocks are qualitatively different. If the wind velocity is higher than the shock velocity, then the shock motion is determined (in the thin shell approximation) by the pressure balance between the wind luminosity at the retarded time and the ram pressure of the external medium. This is the essence of the   \cite{Komp} approximation \cite[see also][] {LaumbachProbstein,ZeldovichRaizer}. In application to winds, the  Kompaneets approximation
 assumes that  the shock dynamics is given by the pressure balance along the shock normal  between the pressure of the radially expanding wind and the ram pressure of the confining medium.

%Interaction of the SN ejecta with the stellar envelope (self-similar solutions \cite{Chavalier82,Nadezhyn85}. 

% \section{Kompaneets-Laumbach-Probstein approximation}

As a model problem,  assume that the  primary SN shock has passed though  the star. As a result,  the density distribution in the expanding envelope will consist of a nearly constant density core and an envelope with a   steep  power law density distribution $\rho\propto r^{-\om}$ \citep{Chevalier82,Nadezhyn85,TrueloveMcKee}; we assume     $\om=9$ for definiteness.  Soon after the passing of the SN shock, a central GBR engine turns on,  acting as an  energy source which with  luminosity  $L_{\rm iso}(\theta)$ constant in time, but depending on the polar angle $\theta$ (the flow is assumed to be axially symmetric). We assume that luminosity is produced in a form of highly supersonic wind with  the velocity close to the speed of light, $v_w \sim c$. The central engine will drive a second shock, which, generally, will be non-spherical.
 We  describe the dynamics of this second non-spherical shock driven by the GBR-type engine into an   external medium with {\it  spherically symmetric } power law density distribution. 
 We  treat the GRB shock dynamics in the  Kompaneets-Laumbach-Probstein approximation, appropriate to  describe accelerating shocks driven by the central source.

 Consider  a small section  of non-spherical 
 non-relativistically
expanding 
contact discontinuity (CD) with radius  $R(t, \theta)$.
 The CD   expands under the ram pressure of the wind,
so that in the thin shell approximation  the normal  stress at the bubble surface is balanced
by the ram pressure of the surrounding medium.
At the  spherical polar
angle $\theta$ the CD  propagates  at an angle 
$
\tan \alpha = - { \partial \ln R / \partial \theta}
$
to the radius vector.
 Balancing the pressure inside the bubble, $\sim L_{\rm iso}/(4 \pi r^2 c)$, 
% (we assume that shock in non-relativistic, while the wind produced by the central source is relativistic), 
with the ram pressure of the shocked plasma, $\sim \rho (\dot{r}- v_0)^2$,  gives \citep[see also][]{Komp,Icke} 
\be
\label{bubbleexp}
\left({\partial r\over\partial t} - v_0\right)^2=  {L_{\rm iso}(\theta) \over4\pi r^2
\rho  c}\left[1+\left({\partial\ln r\over\partial\theta}\right)^2\right]
\label{Komp}
\ee
where $v_0$ is the velocity of radial motion of the   outside medium. We assume highly supersonic flows and neglect for simplicity  the difference between the ram and the post-shock thermal pressures
Below we refer to Eq. (\ref{Komp}) as  the Kompaneets equation. It  describes the evolution of non-spherical shocks with energy supply.

The Kompaneets equation (\ref{Komp}) shows that non-sphericity of the expanding shock   at a given moment  depends   both on 
 the anisotropic driving ($L_{\rm iso}(\theta)$ term)  {\it and} 
collimating effects of the stellar 
material  - the term in parenthesis, which  under certain conditions
tends to 
amplify  non-sphericity. (We assume that  the density  distribution is spherically symmetric). 
Most importantly, Eq. (\ref{Komp}) is non-linear and under certain conditions even a small anisotropy driven by either  the luminosity of the central source or by the initial non-spherical shape of the shock can be amplified and may lead to formation of highly collimated jets. This is indeed what happens in a steep density  part of the post-SN shock profile of  a star. 

\section{Dynamics of secondary GRB shock}

\subsection{Structure of young SNRs }

Interaction of the SN ejecta with the stellar envelope have been considered by \cite{Chevalier82,Nadezhyn85,TrueloveMcKee}. 
After the SN shock passed through the progenitor star, it creates an expanding  SNR with expansion velocity linearly increasing with distance, 
\be
v= { r \over r_0} v_0 = {r \over t}
\ee
where $r_0 = v_0 t$ is the outer radius of ejecta
freely expanding with velocity $v_0$.  
The density structure consists of a  nearly constant density core, and an envelope with a steep density profile created by the SN shock breakout from the surface of the progenitor.
If we assume self-similar expansion, so that at each moment the relative size of the core remains constant, $r_c = \eta_c r_0= \eta_c  v_0 t$, and that the
envelope density profile is a  power law $\rho \propto r^{-\om}$, the density at time $t$ is
\ba && 
\rho = {  f_0 \over t^3}  g(r)
\nn &&
g(r)= \times \left\{
\begin{array} {ll}
1 , \, & r< r_c  \\
 \left({r_c \over r} \right)^\om, & \, r>  r_c 
  \end{array}
  \right\}
  \nn &&
  f_0 = { 3 (\om -3) \over 4  \pi \om \eta_c^3} {M_{ej} \over v_0^3} 
  \label{rho}
  \ea
  The wind  has a fraction $ 3/\om$  of total mass and a  fraction $ 5/\om$ of the total energy. Thus, for $\om=9$ - the fiducial value we will use below - the wind has approximately $30\%$ of mass of the ejecta.

\subsection{Spherical secondary shock}
Let us discuss first the dynamic of a spherical GRB shock propagating through a newly cerated  SNR. In the Kompaneets approximation
\ba && 
{\partial r\over\partial t} = {r \over t} +{ {\cal R } ^{1/4} \over\sqrt{  g(r)}  } {t^{3/2} \over r}
\nn && 
{\cal R } =  \left({L_{\rm iso} \over 4\pi c f_0 }  \right)^2
\ea 
Assuming that the start of the GRB engine is delayed with respect to the SN explosion by time $\Delta t$ \footnote{We assume that the GRB shock  was initiated  when the density structure was given by Eq. (\ref{rho}) with time $t=\Delta t$. Since relations  (\ref{rho}) are asymptotic, for times much longer than the SN shock breakout time, the relations below can be used only as order of magnitude estimates for times
$\Delta t$ of the order  or longer than the SN breakout time.}
and that luminosity $L_{\rm iso}$ is constant in time, 
in the core  (where $g(r)=1$)  the GRB shock propagates according to
\be
r  = 2  {\cal R } ( t+\Delta t) \sqrt{ \sqrt{t+\Delta t} -\sqrt{\Delta t}}=
\left\{
\begin{array}{ll}
\sqrt{2}  {\cal R } \Delta t^{3/4} \sqrt{t}, & t \ll \Delta t
\\ 
2  {\cal R } t^{5/4} , & t  \gg 
\Delta t
\end{array}
\right.
\ee
(GRB central engine starts operating at $t=0$). Thus,  if there is a delay of the switching on of the GBR engine by time $\Delta t$, the resulting 
 GRB shock slows down for a time   $\Delta t$, and only later starts to accelerate.

The GRB shock reaches the edge of the expanding  core  at time and radius
\ba &&
t_{c}  = \left( \sqrt{\Delta t} + \sqrt{t_0} \right)^2 - \Delta t \approx 
\left\{ 
\begin{array}{ll}
2 \sqrt{ \Delta t t_0}   & t_0 \ll \Delta t
\\ 
t_0, & t_0 \ll \Delta t
\end{array}
\right.
\nn &&
R_c = \eta_c  \left( \left( \sqrt{\Delta t} + \sqrt{t_0} \right)^2 - \Delta t\right) v_0
\ea
where 
\be
t_0=    {3 \over 16} \eta_c { M_{ej} v_0 c \over L_{\rm iso}}
\label{t0}
\ee
The time  $t_c$  (or $t_0$) and the radius $R_c$ give  the typical scales of the problem.
In case of time delay between the SN explosion and the activation of the GRB engine, the time it takes for the GRB shock to reach the edge of the expanding core (this time is the main constraint on the model) is the geometrical mean of the dynamical time (\ref{t0}) and the delay time $\Delta t$.

The intrinsic dynamical time $t_0$  and distance at which the GRB shock reaches the surface of the core (for zero delay, $\Delta t=0$) is 
\ba &&
t_0 = 40 \,  {\rm sec} \left({ M \over 2 M_\odot }  \right) \left({ \eta_c \over .3 } \right)  \left({ v \over 10^4 {\rm km s}^{-1}  } \right) 
  \left({ L_{\rm iso} \over  10^{51} {\rm erg s} ^{-1} } \right)  ^{-1} 
  \nn &&
  R_0 = \eta_ c v_0 t_0 =  10^{10} \,  {\rm cm}\left({ M \over 2 M_\odot }  \right) \left({ \eta_c \over .3 } \right)^2  \left({ v \over 10^4 {\rm km s}^{-1}  } \right) ^2
  \left({ L_{\rm iso} \over  10^{51} {\rm erg s} ^{-1} } \right)  ^{-1} 
  \label{tc}
  \ea
  These typical  time and scales are characteristic of the Long GRBs. For a delay of  $\Delta t=50$ seconds, 
  $t_c \sim 200 $ sec and  $R_c \sim 5 \times 10^{10}$ cm. 
    Variations of the GRB engine power, various delays between the activation of the central engine and other variations of the parameters may change these estimates by an order of magnitude in each direction.

\begin{figure}[h!]
\includegraphics[width=1\linewidth]{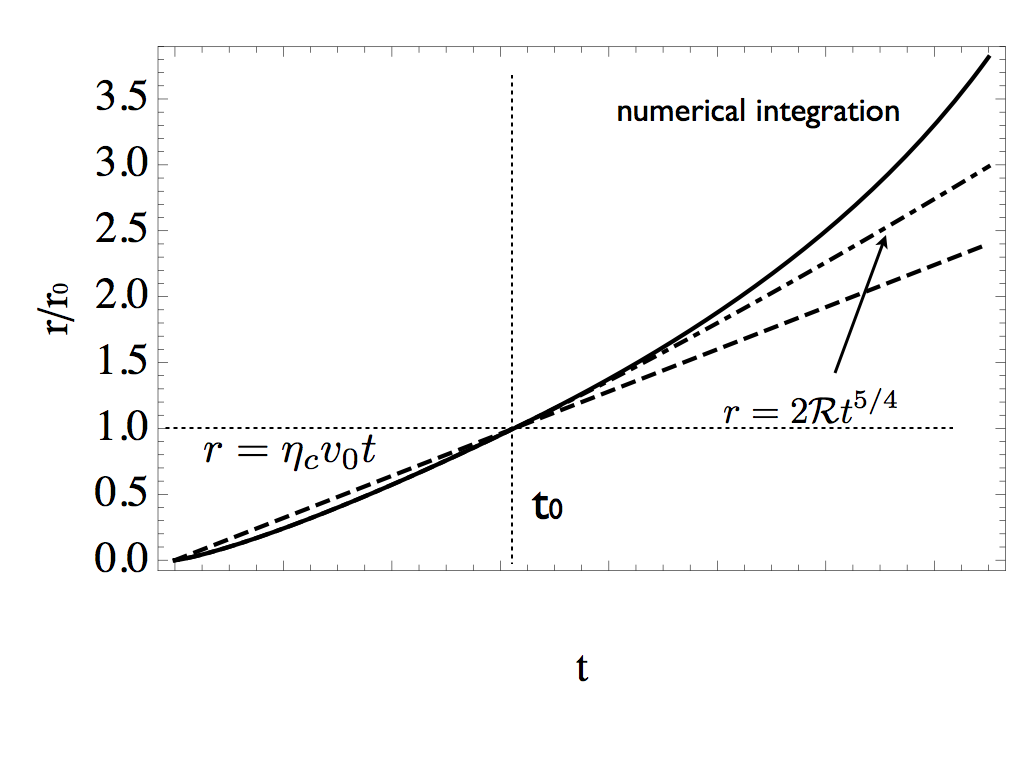}
\vskip -.5 truein
\caption{Comparison of the numerical integration and analytical solutions for spherical expansion of the GRB shock. Shown are the world lines for the surface of the core $r_0 = \eta _c v_0 t$ (dashed) and for the GRB-driven  shock propagating  into expanding SN ejecta. The core boundary is crossed at time $t_0$ when its radius is $r_0 = \eta _c v_0 t_0$. No time delay $\Delta t=0$. In the core,  the numerical  solution follow the analytical result for a   slightly accelerating shock $r \propto t^{5/4}$ (dot-dashed line), while in the envelope it quickly reaches finite time singularity.}
\label{isotropic}
\end{figure}

After the GRB shock reached the  edge of  the core, 
it  moves, approximately for $\om \gg 1$, according to 
\be
r \approx  \eta_c  \left(  1 - { \om -4\over \sqrt{3}}  \left( \sqrt{t\over t_c} -1 \right) \right)^{-2 /(\om-4)}v_0 t
\ee
 For $\om > 4$ the GRB shock experiences finite time singularity, when its velocity formally becomes infinite in a finite time. This occurs at times only slightly longer than $t_c$, by a factor $\sim 1+ 1/\om$, see Fig. \ref{isotropic}.

In what follows,  we present results of calculations for $\om=9$. Results for other power law profiles with  $\om \gg 4$  are similar, while for $\om \rightarrow 4$ the GRB shock break out occurs at very long times.  

\section{Jet formation in the steep density gradient}

Accelerating shocks are typycall  RT unstable \citep[\eg][]{ZeldovichRaizer}. A particularly strong instability develops for shocks that accelerate to arbitrary large velocities in finite time. In such shocks a small perturbation from the spherical shape gets infinitely amplified on the time scale of the acceleration. 

To demonstrate the instability, we  
 numerically  solve Eq. (\ref{Komp}) with density profile given by Eq. (\ref{rho}). We separate the two effects that may contribute to generation of jets -- non-spherical driving and  non-spherical initial conditions. To facilitate comparison with the model results, we introduce new radial coordinate
$r= 2 \tilde{r} t^{5/4}$.   For an isotropic shock inside the core the new coordinate $\tilde{r} $ remains constant, $\tilde{r} = {\cal R } $. 

\begin{figure}[h!]
 \vskip -1 truein
\includegraphics[width=1\linewidth]{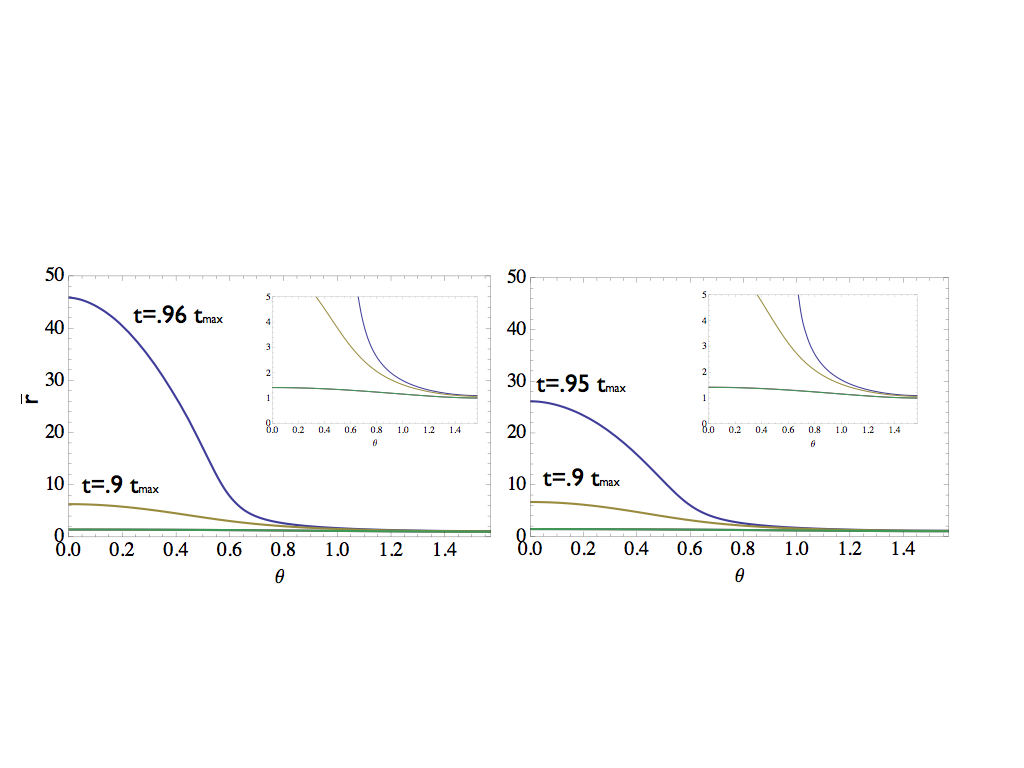}
\vskip -1.25 truein
\caption{Shapes of the secondary GRB-driven  shock in normalized coordinate 
 $\tilde{r}(\theta,t)$. Inserts show that  deep in the core the shape of the shock, $\propto 1 +\cos^ 2\theta$, remains constant and reflects either the weakly anisotropic driving or initial conditions. After the shock reaches the envelope, it develops a  strongly anisotropic shape, producing a highly collimated ''chimney'' in a short time just before the breakout.  {\it Left panel}: anisotropic driver $L \propto  1 +\cos^ 2\theta$, isotropic initial conditions, $r(t=0)=$constant. {\it Right panel}:  isotropic driver $L=$constant, anisotropic  initial conditions, $r(t=0)\propto 1 +\cos^ 2\theta$.}
\label{shocks-shape-1}
\end{figure}

\begin{figure}[h!]
\includegraphics[width=1.4\linewidth]{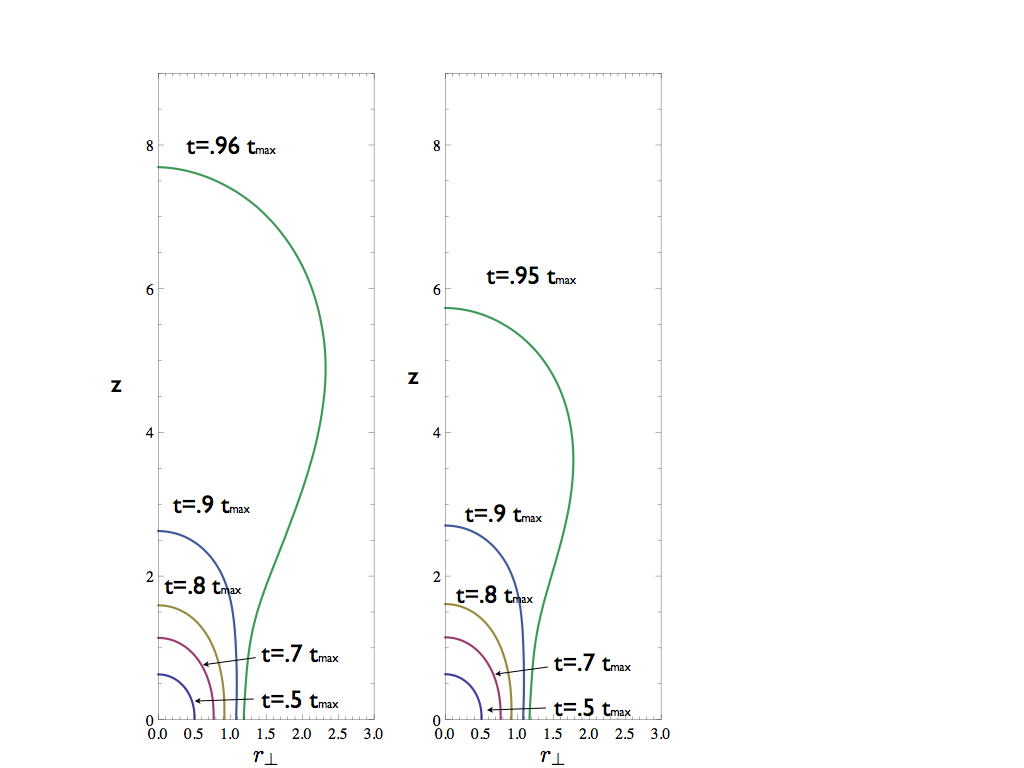}
\caption{Same as Fig. \ref{shocks-shape-1}, but in terms of the
physical distance ${\bf r(t)} $. }
\label{shocks-shape}
\end{figure}

First, we consider a shell of spherical form   at $t=0$ located at small distance  and  driven by non-spherical wind. As a simple exemplary case we chose wind luminosity  $f(\theta)= 1+\cos^2 \theta$, so that the wind power along the polar direction   is two times of that along the equator. 
Numerical integration shows that   at times $t\leq t_0$, the evolution proceeds in a linear regime, so that  a weakly anisotropic driving results in a weakly anisotropic shock.
Soon after the shock reaches the edge of the core  at $t_0$ and enters the steep density gradient in the envelope, it becomes unstable and develops  
 a collimated jet, see Figs. \ref{shocks-shape-1}- \ref{shocks-shape}. For $\om \gg 4$, the jets develops on the time scale $t_0$.
 
 Secondly, we  consider an isotropic driver, $f=1$ but starting with a non-spherical initial shape, $r(t=0)= 1 + \cos^2 \theta$. In this case the GRB shock dynamics is qualitatively similar to the non-spherical wind: the shock remains weakly anisotropic in the core and develops a jet soon after reaching the envelope, Figs. \ref{shocks-shape-1}- \ref{shocks-shape}. 
 
\section{Discussion}

We suggest that the outflows driven by  core-collapse supernova  generically have two driving mechanisms: neutrinos and a GRB-like central engine. 
Most supernovae are dominated by the neutrino-driven quasi-spherical outflow,  while long  GRBs are dominated by the central engine, producing relativistic collimated jets.  The second GRB-driven shock propagates into a density structure created by the primary SN shock. This density structure of early SNRs has a central core with nearly constant density and an envelope with  steeply decreasing density.  In the envelope the 
 secondary  shock is accelerating,  is unstable to  Raleigh-Taylor instabilities and may form ``chimneys'', cylindrically collimated outflows. Jet formation occurs even if the central engine produces only weakly anisotropic wind. This is  a non-linear processes, by which a non-sphericity of the shock is amplified by the wind. The formation of  the ``chimney''  occurs over extremely short dynamical time, while  the time of  the ``chimney'' formation is related to break out time. It is also required that the GRB engine is sufficiently powerful, so it can drive the secondary GRB shock  through the constant density core on sufficiently short time scales. 
 Thus, there is a lower power threshold  on the GRB engine required for the production of collimated outflows. We suggest that if this threshold is not reached, weakly anisotropic SNe are generated; \eg, Cas A  and SN 2009bb may be examples of such failed GRBs.

We presented a numerical model illustrating the formation of collimated outflows in a model expanding star. The model is naturally idealized, we assumed the SN ejecta structure is described by the  asymptotic profile, which takes a SN shock break out time to establish, of the order of hundred seconds, which is similar to the  dynamical time (\ref{tc}). 

%K.M. Schure1, J. Vink1, A. Achterberg1andR. Keppens1,2,3: they go until R =10^{18} cm.

\bibliographystyle{apj}
\bibliography{/Users/maxim/Home/Research/BibTex}

 \end{document}